# A Detailed Numerical Analysis of Asymmetrical Density Distribution in Saturn's F ring During an Encounter with Prometheus

Phil J. Sutton*

p.j.sutton@lboro.ac.uk

+44(0)1509 304333

Feo V. Kusmartsev

f.kusmartsev@lboro.ac.uk

 +44(0)1509 223316

Astronomy Unit, Physics department, Loughborough University, Loughborough, Leicestershire, LE11 3TU, UK

Key words: *gravitation, planets and satellites: rings, protoplanetary discs, planets and satellites: formation*

# ABSTRACT

Saturn's rings, reminiscent of an early Solar System present a unique opportunity to investigate experimentally some mechanisms thought to be responsible for planet and planetesimal formation in protoplanetary discs. Here we extended the comparison of our numerical models of Prometheus encountering the F ring employing non-interacting and interacting particles. Higher resolution analysis revealed that the density increases known to exist at channel edges is more complex and localised than previously thought. Asymmetry between density increases on channel edges revealed that the channel edge facing way from Prometheus to be the most stable but with lowest maximum increases. However, on the channel edge facing Prometheus the interacting model showed large chaotic fluctuations in the maximum density of some clumps, much larger than those of the other channel. The likely cause of this asymmetry is a variance in localised turbulence introduced into the F ring by Prometheus. High resolution velocity dispersion maps showed that there was a spatial link between the highest densities and the highest velocity dispersions in the interacting model. Thus suggesting that the high velocity dispersion we see is the reason for the observed inhomogeneous distribution of fans (evidence of embedded moonlets) on some of the channel edges facing Prometheus.

# 1. INTRODUCTION

Saturn's F ring, a unique place in the Solar System, has been the subject of intense studies due to the diverse dynamics witnessed by the CASSINI spacecraft over short time frames. The dusty F ring takes on a knotted asymmetrical structure where most of its mass is assumed to be in its central core. Low density spiral strands have been observed to reside either side of the core, their formation arises from the long term evolution of jets by physical collisions of small moonlets on eccentric orbits through the central core (Charnoz et al 2005, Murray et al 2008). The central core is known to be home to a large population of resident small moonlets from stellar occultations (Meinke et al 2012; Meinke et al 2011; Hedman et al 2011), the presence of mini jets (Attree et al 2013; Attree et al 2012) and fan structures emanating out from the core (Beurle et al 2010). Interactions between Prometheus, the F rings inner Shepherd moon, and the F ring is thought to be well understood with many of the features, structures and the large population of transient moonlets directly attributed to Prometheus (Sutton & Kusmartsev 2013; Beurle et al 2010; Murray et al 2008; Murray et al 2005). However large amounts of small moonlets responsible for creating mini jets in the central core appear to have no direct link to Prometheus in their position (Attree et al 2013), raising more questions about their origin.

## 1.1 Moonlet formation

Some elements of moonlet formation in the F ring was found to be directly linked to the perturbations of Prometheus, with fans, the structural signatures of moonlets spaced at the same 3.27° as Prometheus induced structures in CASSINI images (Beurle et al 2010). A numerical study assuming collisionless non-interacting particles within periodic boundary conditions was used to simulate the local effects of density fluctuations on a near homogeneous ring post Prometheus encounter. It was discovered that density increased at channel edges up to a maximum 2.5 times the original density, with large fluctuations in the maximum number density of particles occurring over one orbital period. These large variations in local density can be seen to relate inversely to the magnitude of the minimum radial velocity. This then revealed that maximum number densities occurred when the lowest minimum radial velocities were seen. A clear link to moonlet formation and Prometheus in the F ring was provided but left open a window for more detailed investigations into the density evolution during an encounter with Prometheus, which is where our model seeks to expand.

A predator – prey model was proposed by Esposito et al 2012 for the edge of B ring and the F ring. Here aggregate size was seen as the prey and the perturbing moons as the predator. It was shown that Prometheus has an influence on clump formation in the F ring with Mimas displaying a comparable dominant effect on the outer B ring edge. Locations of moonlets or clumps were seen to be at resonances in the rings with the predator moons. Large increases in velocity dispersion were seen to decrease the local density with disaggregation developing from disruptive collisions or tidal shear. However agitation of ring material at the ring edge through stochastic processes can yield the increase of density of clumps into more persistent objects within the ring, again forming a link to possible moonlet formation and the perturbing moon.

Another mechanism was suggested for moon and moonlet formation by viscous spreading to explain the current configuration of Saturn's moons and moonlets (Charnoz et al 2010). Here the main moons and moonlets both showed a linear increase in mass with distances from Saturn but with distinctly different gradients. The differences, when elaborated established the moonlet group to exhibit a larger mass enhancement with increasing distances from Saturn in comparison to the main moons. Moonlets growth at the rings edge is attributable to positive induced torque from the rings and the planet driving their outward migration. As the torque ($\Gamma_s$) increases with the moon mass ($m_s$) $\Gamma_s \propto m_s^2$ the migration rate also increases, driving the faster outward migration of more massive moons. Due to the different migration rates, orbital crossings and merging can occur which leads to the arrangement of moons radially aligned as a function of their mass. This mechanism can therefore explain the existence of dynamically young moons so close to the rings.

## 1.2   Structures From Embedded Moonlets

Propeller shaped structures, the signature of small radially stationary embedded moonlets a few hundred meters in size, have been observed in Saturn's rings by CASSINI (Tiscareno et. al., 2008, Srem evi et al. 2007, Tiscareno et al. 2006). Previous numerical work used different methods to our model where physical collisions, particles sizes and periodic boundary conditions were all implicit in their calculations (Lewis & Stewart 2009). Here in the A ring, where particle sizes and surface densities are at the highest, physical collisions produce a relative dampening effect on structures created by moon / moonlet interactions. This case is shown in greater detail where propeller shaped structures formed by stationary moonlets in the rings were investigated. When non-interacting test particles were used a repeating pattern of the particles trajectories with a periodicity of $3\pi\Delta a$ where $\Delta a$ is the orbit separation was seen. However when collisions and self-gravity was taken into account a damping effect was seen where eccentricities were reduced and orbits randomised beyond the initial structure. It is likely that physical collisions would have a similar effect on other types of structure created by embedded moonlets or perturbing moons. Consequently in rings that have smaller particles sizes, F, G, E rings, and much lower surface densities physical collisions would have a less dominant outcome in the evolution of moonlet induced structures. Instead inter-particle dynamics are dominated by gravitational forces.

## 1.3   Shepherded Debris Disks

Saturn's rings role as a local laboratory to study processes and evolution of larger scale astrophysical discs is an important one and the most dynamic of them all, the F ring, could prove to be very useful in the study of narrow shepherded debris discs. For example the Fomalhaut system is a narrow debris disk orbiting its host star; reminiscent of Saturn's F ring it owes its narrow structure to two theorised shepherding planets (Boley et al 2012 a). Figure 1 shows a comparison of the two systems. The sharply truncated inner edge of the ring and eccentricity are good indicators of an internal shepherding planet that could be responsible for shaping it in the same way Prometheus dominates the sculpture of Saturn's F ring. It is still in debate whether Fomalhaut b is responsible for the inner shepherding of the disk or a pair of currently unseen shepherding planets is the cause of the narrow structure. However recent work by ALMA (350GHz) failed to observe any shepherding planets suggesting a much smaller size for the rings creators than previously thought (Boley et al 2012 b). Evidence that the possible cause of the rings morphology is the existence of two small shepherding planets was still supported. Again constraining the idea that it could resemble a similar setup to Saturn's F ring where small moons shepherd but don't destroy the F ring. Using values taken from numerical studies by Chaing et el 2009 it is possible to make assumptions that the inner planet could experience close encounters with the debris disk similar to Prometheus. For this we need to assume that both the disk and inner shepherd planet are on elliptical orbits (0.11 and 0.12 eccentricity respectively) and that there is some degree of mutual precession of their orbits due to a non-spherical host star. If this is the case its appearance can be assumed to be similar to the F ring. With semi major axis of 115 AU for the inner planet and 133 – 158 AU for the disk at anti-alignment and closest approach we would see the inner planet at apoapsis and the disk at periapsis. This would be at a radial distance from Fomalhaut of 128.8 AU for the inner planet and 118.37 – 140.62 AU for the disk. This would lead to a very big disturbance of the disk due to the planet. However the rate of precession or any more details about the likely precession are difficult to calculate due to the uncertainty of the Fomalhaut system.

Future observations of the Fomalhaut debris disk with the proposed EPICS detector on the EELT could help give higher resolution investigations of the disk. The science goals of the EPICS detector are a resolution of 0.005" with a field of view 1.37" x 1.37" (Kasper 2008).

Assuming these and applying it to the Fomalhaut debris disk in question, we can approximate that the field of view would cover approximately 30AU x 30 AU. If we make another assumption that any structures formed by an inner shepherding planet would be separated by the same as Prometheus induced structures in the F ring then there will be a spacing of approximately 7AU between areas of high and low density or surface brightness. This is well within reach of the EPICS detector, although will not be anywhere near the same resolution as CASSINI can currently achieve. It should still be able to detect any asymmetry created by closely interacting shepherding planets. It is likely that the larger separation and eccentricities between disk and planet compared with the F ring and Prometheus would lead to a larger difference in orbital periods and thus a larger separation in structures. A separation of ~ 10 – 20 times that of streamer-channels in the F ring is possible, which again should be within reach of the EPICS resolution.

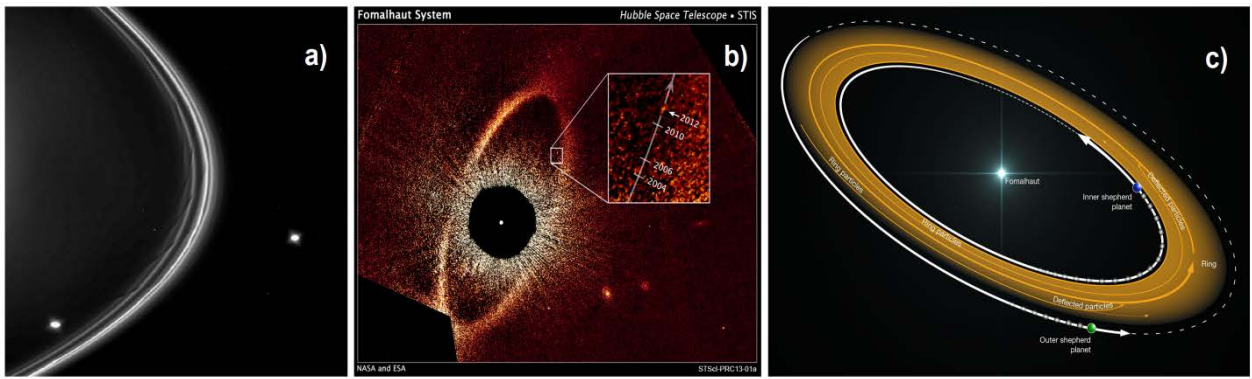

**Figure 1 |** a) An image taken by CASSINI of Saturn's F ring, shepherded by its two moons Prometheus (inner) and Pandora (outer), b) image taken by the Hubble Space Telescope showing the debris disk around Fomalhaut with the zoomed in section illustrating the movement of the inner shepherding planet (Fomalhaut b) over the course of 8 years and c) an visual representation of the Fomalhaut system depicting the two shepherding planets that could be responsible for the narrow ring, similar to Saturn's F ring. Image credit: ALMA (ESO/NAOJ/NRAO)/B. Saxton. All of the images are orientated to show an anti-clockwise orbital direction.

## 2. NUMERICAL METHOD

To obtain the results reported in our manuscript we employed an n-body simulation of 6.5 x$10^6$ point like particles where they evolved in the system due to gravitational forces taken from classical Newtonian mechanics only (Springel 2005). Particle surface densities of a maximum of 0.17 per $km^2$ are similar in value to previous numerical studies of the F ring (Chavez et al 2009; Murray et al 2005). Particle collisions in their models showed very few ring particles collided with Prometheus, ruling out physical collisions as the cause of the streamer-channels. Thus neglecting physical collisions in our models has negligible effect on the initial Prometheus encounter. Equations of motion for each particle as well as for Prometheus and Saturn can be shown as:

$$\ddot{\boldsymbol{r}}_s = -G \sum_{i=1}^{N}(\boldsymbol{r}_i - \boldsymbol{r}_s)\frac{M_i}{|r_i - r_s|^3} - G(\boldsymbol{r}_p - \boldsymbol{r}_s)\frac{M_p}{|r_p - r_s|^3} \qquad [1]$$

$$\ddot{\boldsymbol{r}}_p = -G \sum_{i=1}^{N}(\boldsymbol{r}_i - \boldsymbol{r}_p)\frac{M_i}{|r_i - r_p|^3} - G(\boldsymbol{r}_s - \boldsymbol{r}_p)\frac{M_s}{|r_s - r_p|^3} \qquad [2]$$

$$\ddot{\boldsymbol{r}}_i = G \sum_{j \neq i}^{N}(\boldsymbol{r}_i - \boldsymbol{r}_j)\frac{M_j}{|r_i - r_j|^3} - G(\boldsymbol{r}_p - \boldsymbol{r}_i)\frac{M_p}{|r_p - r_i|^3} - G(\boldsymbol{r}_s - \boldsymbol{r}_i)\frac{M_s}{|r_s - r_i|^3} \qquad [3]$$

Where all of the vectors in the above equations are taken from Saturn also located at the origin. Prometheus was assumed to be at the periapsis of its orbit at the start. Two separate numerical models were created where the only difference was the mass of the ring particles. The first model assumed all ring particles to be massless test particles moving only under the influence of Saturn and Prometheus. The second model assumed a single ring particle mass of 10kg whose motion is influenced not only by Saturn and Prometheus but also every other particle in the F ring. In the case of this manuscript both models were allowed to evolve for a longer period of time than previously where the initial encounter of Prometheus was investigated (Sutton & Kusmartsev 2013).

An integration method was used to reduce the overall computation time down significantly by using a TreePM code parallelised across multiple processing cores by means of Peano-Hilbert domain decomposition. To reduce the overall number of force calculations close range forces were computed using a Barnes–Hut Tree code whilst long range forces used a PM (particle mesh) method. Where the Particles involved in long range force calculations were grouped together with other nearby particles in a clouds-in-cells (CIC) approach with the forces calculated using a FFTW technique (Springel 2005). In the TreePM method a discretized particle system is mapped into continuous model with the peculiar potential defied as:

$$\emptyset(x) = \sum m_i \varphi(x - x_i) \qquad [4]$$

Where $\varphi(x - x_i)$, a single particle gravitational potential, is used to get gravitational forces in the equation of motion above, in eqs (1-3). Then this potential, eq. (4), is split in Fourier space into a long-range and short-range part according to $\emptyset_k = \emptyset_k^{long} + \emptyset_k^{short}$, where:

$$\emptyset_k^{long} = \emptyset_k \exp(-k^2 r^2) \qquad [5]$$

$$\emptyset_k^{short} = -G \sum \frac{m_i}{r_i} erfc\left(\frac{r^i}{2r_s}\right) \qquad [6]$$

Where $r_s$ is the spatial scale of the force split,

Adaptive time stepping was also used for all particles where time steps of particles were integer values of one another. This reduced down the overall forces computed, with particles time steps derived based upon its acceleration and assigned smoothing length (Springel 2005). The time step of each particle can then be shown as:

$$\Delta T_{grav} = min\left[\Delta t_{max}, \left(\frac{2\eta\varepsilon}{|\alpha|}\right)^{1/2}\right] \qquad [7]$$

Where $\Delta t_{max}$ is the maximum allowed time step,

$\alpha$ is the acceleration of the particle,

η is the accuracy parameter,

$\varepsilon$ is the smoothing length of the particle

No hydro-dynamical forces were included within the calculations, only gravitational forces. We also did not include direct physical collisions between particles instead two particles interacting have a reduced (or smoothing) gravitational force once the distance between them became smaller than some characteristic smooth length common for all ring particles. Saturn and Prometheus have different, larger smoothing lengths, associated with their masses. In contrast to protoplanetary discs pressure is not critical when considering Saturn's rings, and therefore the approximation of gravitational forces only is well suited for this case. Spatial boundaries on the system have not been defined and so the system was open with particles allowed to evolve in free space. This is hoped to reduce any errors that could be introduced as a result of using a shearing box approximation on the F ring and periodic, quasi-periodic and chaotic structures formed by Prometheus.

## 2.1   Saturn's F ring

The starting position of particles was derived from parameters previously used for numerical modelling of the F ring (Murray et al 2005; Beurle et al 2010; Sutton & Kusmartsev 2013), strands and the central core. To create a more realistic F ring structure all ring particles were arranged randomly into four groups or rings around Saturn, with the first being a background sheet of particles and the subsequent three groups being the inner strand, central core and outer strand respectively. The central core contained $2 \times 10^6$ particles to account for the suspected higher particle densities that are present in the core. The inner strand, outer strand and background population group all contained $1.5 \times 10^6$ particles each, distributed randomly with equal probability around the whole ring. The strands and core were all assumed not to be spiral in nature for the sake of our modelling where the true trajectories of particles within these strands can be difficult to model. These distributions are associated with the initial conditions of the particles used in our numerical modelling. They automatically give rise to higher particle number densities in the inner and outer strand and higher again in the central core. This choice has been based on the observations made by CASSINI, which have suggested a higher density in a central core and strands (assuming a higher surface brightness). However, when considering the initial conditions for our numerical model, due to the likely chaotic and non-uniform distribution of clumps or moonlets located in the core (Attree et al 2013) we neglected their presence. It should then be important to note that most of the mass in the central core and the whole F ring is thought to be confined in the large population of moonlets in the core, which we have omitted from our simulations (Scharringhausen & Nicholson 2013). The existence of these objects in the core would certainly have an effect on the local density changes during an encounter of Prometheus, but it is also probable that some of these same objects are the consequence of such encounters. Therefore by studying the relative changes in density as Prometheus disrupts the F ring we hope to further understand the origin of the randomly distributed moonlets. A reference frame has been chosen with Saturn placed at the origin of our system of coordinates, where its initial conditions are associated with a zero magnitude velocity vector. Therefore the equations for initial positions of all particles can be shown as:

$$\boldsymbol{R_s} = [0,\ 0,\ 0] \qquad [8]$$

$$\boldsymbol{R_p} = [139{,}671\ km,\ 0,\ 0] \qquad [9]$$

$$\boldsymbol{R_j} = [r*\cos\theta,\ r*\sin\theta, 0] \qquad [10]$$

Where $r$ represents the radial position of ring particles from Saturn and $\theta$ the angular position of ring particles around Saturn. All ring particles are assumed to have circular-like trajectories located within the F-ring; this has been done to help with creating comparable figures that have multiple stages of evolution with respect to orbital periods since the beginning of the numerical model. The initial positions for the radial distances of particles from Saturn are split into four groups or rings to represent the background sheet of particles, inner strand, central core and outer strand respectively.

$$r(0{:}1{,}499{,}998) = r_1 + w_1 * random(n_1) \qquad [11]$$

$$r(1{,}499{,}999{:}3{,}499{,}998) = r_2 + w_2 * random(n_2) \qquad [12]$$

$$r(3,499,999: 4,999,998) = r_3 + w_3 * random(n_1) \qquad [13]$$

$$r(4,499,998: 6,499,999) = r_4 + w_4 * random(n_1) \qquad [14]$$

Where the number inside the brackets represents the particle ID belonging to each of the four groups. The values $r_1, r_2, r_3, r_4$ are the radial distances from Saturn to the inner ring boundary for each of the four ring groups, $w_1, w_2, w_3, w_4$ are the widths of each ring and the function *"random(n)"* represents a random number generated from 0.000 to 0.999 $n_1$(1,500,000) or $n_2$ (2,000,000) times.

Each ring inner boundary and width can be shown as:

$r_1 = 139,876 \, km \qquad w_1 = 700 \, km$

$r_2 = 140,049 \, km \qquad w_2 = 70 \, km$

$r_3 = 140,214 \, km \qquad w_3 = 20 \, km$

$r_4 = 140,299 \, km \qquad w_4 = 30 \, km$

An angular position $\theta$ for each of all ring articles has been taken as:

$$\theta = 2\pi * random(N - 2) \qquad [15]$$

Velocities of all particles in the initial conditions of our numerical modelling have been derived from equations for circular orbits, and given by the following equations:

$$\dot{R}_s = [0, \ 0, \ 0] \qquad [16]$$

$$\dot{R}_p = [0, \ v_p, \ 0] \qquad [17]$$

$$\dot{R}_J = [v * \cos\theta_v, \ v * \sin\theta_v, \ 0] \qquad [18]$$

Where: $\qquad v_p = \sqrt{\dfrac{(G(M_s + M_p))(1+e)}{(1-e)*a}} \qquad [19]$

And Prometheus is assumed to be starting at the periapsis of its orbit. This condition is assumed to be the case throughout the manuscript; the parameters of its orbital are taken from Spitale et al (2006).

The magnitude and angle of the initial velocity vectors of all ring particles can be expressed through the generated random numbers (see, eqs (8-12)) with the use of the formulae:

$$v = \sqrt{\frac{G(M_s + M_j)}{r}} \quad [20]$$

$$\theta_v = \theta + \frac{\pi}{2} \quad [21]$$

## 2.2 Density Analysis

All rendered density plots were created using original snapshot files output from our numerical code and used a fake smoothing length assigned to each particle to create an artificial particle density which could be represented relative to all other particles in the ring. This was then used to visual the relative number density of particles in our figures using SPLASH (Price 2007) a tool for the visualisation of SPH numerical simulations. Only ring particles with the same mass were used in the creation of density plots i.e. only ring particles. Additional programs were also written in IDL and were used to calculate the maximum number density and the average density of particles within the box at specific points in space and time arising at the evolution of the F-ring. Here, unlike previously where a 1000 x 1000km box was placed around channel edges (Sutton & Kusmartsev 2013); we selected the clumps or areas of highest local density and placed a 200 x 200km box around them. By placing a smaller box around the area of interest the particle's number density was measured along with the number of particles in clump, producing a detailed analysis of evolution of density in the clumps. The results obtained with the use of the two models have then been compared in this manuscript.

### 2.3 Rendered Velocity Dispersion Plots

To further investigate the nature of turbulence in the F ring attributed to Prometheus encounters additional analysis was done of the outputted data files. Here particles velocity magnitude had its unperturbed counterpart removed from its actual value. The resultant magnitude is used to create a rendered plot that shows the spatial distribution of changes in particles velocity magnitudes. With this approach a spatial investigation into the velocity dispersion and density enhancement can be accomplished.

### 3. SIMULATION RESULTS

We ran two simultaneous models with and without ring mass (non-interacting and interacting respectively) and analysed in greater detail the density enhancements previously reported (Beurle et al 2010; Sutton & Kusmartsev 2013) at the channel edges caused by the interactions of Prometheus on the F ring. Table 1 and 2 show the Max No. density and average particle density for each clump or where there was an extended area of density enhancement centred over the maximum density.

| Orbital Period (T) | Non-interacting | | Interacting | |
|---|---|---|---|---|
| | Max No. Density | Average particle density | Max No. Density | Average particle density |
| 0 | 13 | 5.47 | 13 | 5.47 |
| 1.5 | 27 | 8.61 | 32 | 9.41 |
| 2.5 | 17 | 6.30 | 19 | 6.34 |
| 3.5 | 26 | 8.72 | 22 | 8.63 |
| 4.5 | 21 | 8.11 | 27 | 8.85 |
| 5.5 | 26 | 10.15 | 26 | 8.59 |
| 6.5 | 27 | 10.16 | 29 | 10.35 |
| 7.5 | 23 | 8.95 | 26 | 10.14 |
| 8.5 | 30 | 9.80 | 31 | 10.09 |
| 9.5 | 27 | 9.22 | 29 | 9.93 |
| 10.5 | 27 | 9.45 | 26 | 10.05 |

**Table 1 |** The Max No. Density of particles and the average particle density within the clumps identified at the channel edge facing away from Prometheus are shown for both the interacting and non-interacting model.

| Orbital Period (T) | Non-interacting | | Interacting | |
| --- | --- | --- | --- | --- |
| | Max No. Density | Average particle density | Max No. Density | Average particle density |
| 0 | 13 | 5.47 | 13 | 5.47 |
| 1.5 | 20 | 7.22 | 15 | 5.66 |
| 2.5 | 16 | 6.30 | 19 | 6.05 |
| 3.5 | 23 | 7.51 | 22 | 7.74 |
| 4.5 | 25 | 8.44 | 24 | 8.61 |
| 5.5 | 27 | 8.21 | 23 | 8.35 |
| 6.5 | 24 | 8.69 | 29 | 11.48 |
| 7.5 | 28 | 9.41 | 25 | 9.17 |
| 8.5 | 23 | 9.20 | 37 | 10.75 |
| 9.5 | 26 | 8.57 | 29 | 9.18 |
| 10.5 | 27 | 10.54 | 28 | 10.14 |

**Table 2 |** The Max No. Density of particles and the average particle density within the clumps identified at the edge facing towards Prometheus are shown for both the interacting and non-interacting model.

### 3.1 Model Comparison With The Real F ring

In our simulations the total mass of the F ring modelled is 4.075 x $10^{-10}$ $M_p$. This is much less than the predicted mass of the clumps in the F ring. If clumps are considered in the central core their inhomogeneous distribution in the initial conditions makes investigating density variations after the Prometheus encounter become difficult to model. Therefore we have chosen to neglect the randomly distributed moonlets known to exist in the central core which make up the majority of the mass in the F ring. Instead focusing on the asymmetry first produced during the initial encounter. Surface densities are then 0.00068 kg m$^2$, 0.0068 kg m$^2$, 0.032 kg m$^2$ and 0.0159 kg m$^2$ for the background sheet of particles, inner strand, central core and outer strand respectively. These are much lower surface densities then those used and assumed for the main rings. There it is assumed that the dusty F, G and H rings have considerably lower surface densities than the A, B and C rings system. The number of particles in each of the components is what then affects the surface density of individual strands / core. Also recent developments into the physical properties of the dusty F ring show particles sizes to be considerably smaller than the major Saturnian rings, ~0.5µm (Scharringhausen & Nicholson 2013) compared with characteristic particle size distribution in the A and B rings of 30cm $a_{min}$ to 20m $a_{max}$ (French & Nicholson, 2000). Discounting the mass of the moonlets located in the central core the mass of the remaining F ring is almost negligible (Scharringhausen & Nicholson 2013).

Our model also assumes a monolayer where in reality the F ring and other rings have some vertical component which also effectively defines the optical depth and subsequent surface density. The F ring actually has the largest vertical component to it, equivalent depth of 10 ± 4 km (Scharringhausen & Nicholson 2013) compared with at least an order of magnitude smaller for the main rings. It should be natural then to assume that future numerical studies of the F ring should include a multi-layer / stratified structure, as the vertical component to particles dispersions should play an important role in the chaotic and turbulent environment of the F ring.

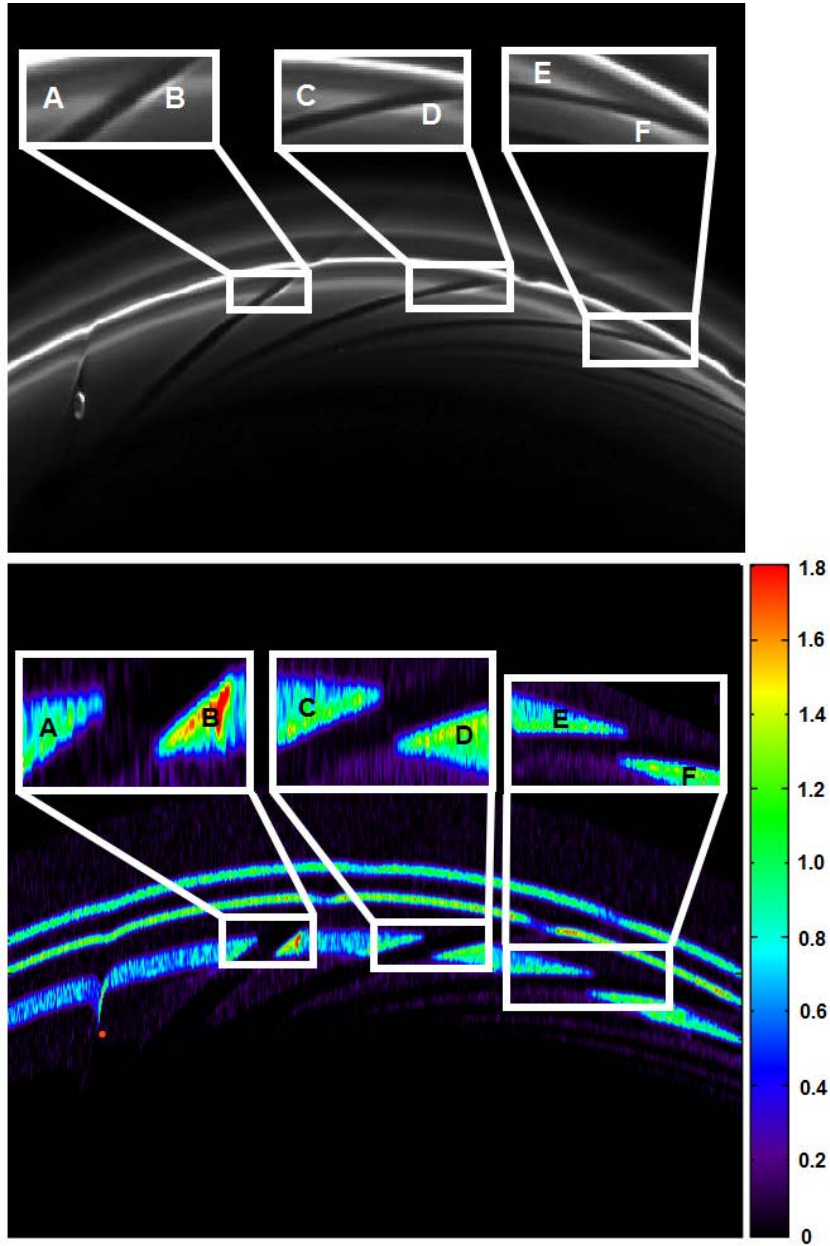

**Figure 2 |** a) An image taken by CASSINI on June 1,' 2010 in the narrow angle camera. During the time of the image CASSINI was 808,000 miles from Saturn relating to a 5 miles per pixel resolution. Here assumptions are made that surface brightness seen in the CASSINI images is proportional to surface density (calculated with our model). Surface brightness is normalised to the inner strand. b) A rendered plot depicting the density calculated with our model at the same orbital phase as seen in the image by CASSINI. This density is normalised so that it has a value equal to 1 on the unperturbed inner strand, i.e. the same as in the CASSINI image, making density changes in the theory (model) and observations more comparable (Table 3, below). Areas A, C, E represent locations are on the

channel edge facing away from Prometheus while areas B, D, F represent locations on the channel edge of the inner strand facing towards Prometheus post encounter. An orbital phase of 0.57 was assumed to match the observations in our model where the apoapsis of the orbit was at 0.5 and periapsis at 0.0 and 1.0. It is also noted that channels appear at their most open during the apoapsis of Prometheus' orbit and thus channel edges will show their highest densities at this point.

| Region | CASSINI (% change in surface brightness) | Model (% change in particle density) |
|---|---|---|
| A | +75% | +83% |
| B | +10% | +64% |
| C | +58% | +70% |
| D | +9.6% | +62% |
| E | +56% | +60% |
| F | +9.7% | +58% |

**Table 3 |** The locations (A, …,F) identified in figure 2 are listed with their derived change in brightness (CASSINI) and density (modelled). Here, for comparison, we have presented the maximum values at these locations.

Due to resolution limitations in images taken by CASSINI comparable quantitative data of the change in density was difficult to obtain. However it can be easily seen that there is a clear asymmetry in structures formed by Prometheus noticed in both our simulations and the real F ring, Fig 2. There is a high discrepancy between the model results and the real F ring. It arises on the channel edge facing away from Prometheus at this time moment. It is because areas of enhanced density on this channel edge obtained in our model are highly localised. It is likely that the resolution in CASSINI images effectively smooth's out highly localised density increases. On the other hand in our simulations we have significantly higher capabilities to resolve genuine maximum increases. In our models the channel edge that faces Prometheus, (B, D, F) a larger area of enhanced density is seen; after 5 orbital periods there arise large chaotic like fluctuations in the maximum density.

We also note that as the system evolves density enhancements become more dominant in the central core, residing in the area where a large population of moonlets have been observed (Meinke et al 2012, Attree et al 2012, and Attree et al 2013). There is then a clear spatial correlation between surface brightness and surface density which occurs during the evolution of various structures.

## 3.2   Localised Density Enhancements

The largest clumps seen at the channel edges were then tracked over multiple orbital periods and compared in Tables 1 and 2. Previously the channel edges were shown as increasing in density over time but our results demonstrate that these increases are more localised. This local increase in density within the channel edges is markedly different between the two models, as can be illustrated in Fig 3.

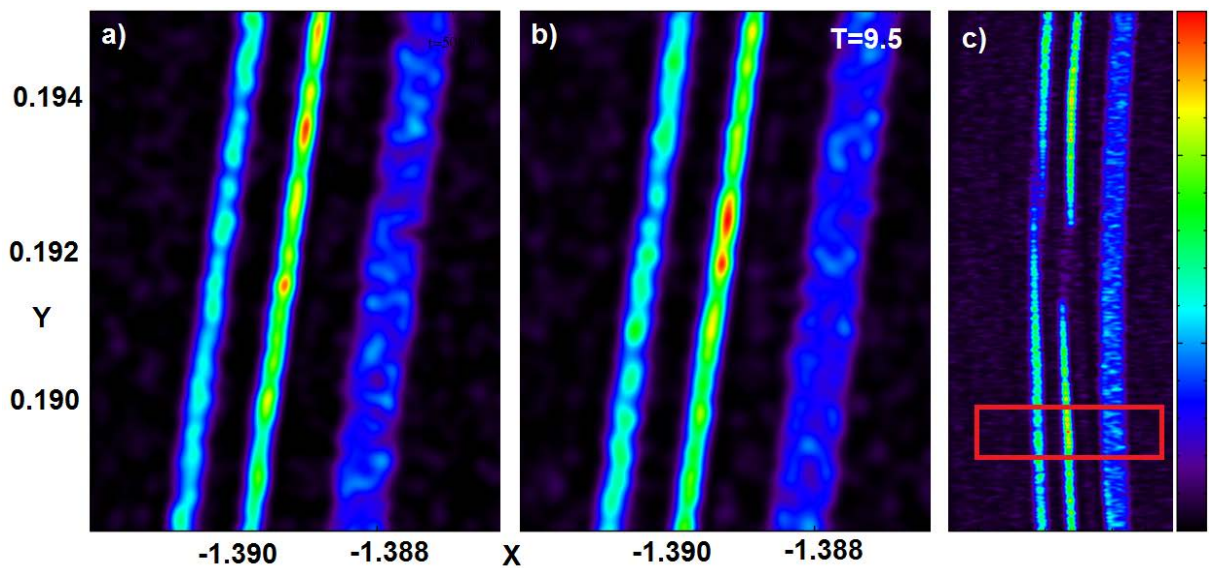

**Figure 3 |** A comparison between the two models is shown at a time of T=9.5, where T is given in Prometheus orbital periods since the start of the simulation and **a)** is the non-interacting model, **b)** is the interacting model and **c)** is a visual representation using the interacting model to show the position on the channel where the snapshots **a** and **b** are taken from. Here the reference frame is zoomed in to show the clump with the highest density on the channel edge facing away from Prometheus. While the non-interacting model shows a larger single clump associated with much more coherent and defined area of increased density. Although the average density and max number density for both areas is only slightly higher for the interacting model it does clearly show a well-defined and larger area than that obtained in the framework of the non-interacting model.

Figure 3 illustrates the distinct composition of the localised increase in density witnessed at one of the channel edges 9.5 orbital periods after the initial encounter. What we see is that the non-interacting model shows multiple detached areas of enhanced density whereas the interacting model displays a single well defined area of increased density. Individual clumps identified in both models do not differ drastically in the Max No. Density and average number density. However it is their shape and ultimate distribution within the channel edge that is in contrast between the models, at 9.5 orbital periods. Although here we only use the channel edge facing away from Prometheus after 9.5 orbital periods it must be noted that a similar trend was seen at the same channel edge throughout their evolution. This difference between the two models became more apparent as the two models were allowed to evolve over a longer time.

Density profiles of clumps seen in Fig 3 show that at the channel edge facing away from Prometheus the average particle number density and the Max No. Density for both models is different (Fig 5b). The interacting model at this point in the evolution of the system exhibits a Max No. Density approximately 7.4% higher than the non-interacting model, while the average particle number density is 7.7% higher than the non-interacting model. In the interacting model a higher proportion of particles with a number density greater than 20 can be seen causing the noticeable difference in average density in the clump. It should be noted here that both models show a similar difference in their maximum and average number densities with very little fluctuations throughout their evolution, thus suggesting a higher degree of stability at this location.

When we consider the same process for the opposite channel edge it is noted that evolution of the density enhancements does not completely mirror the other edge. Figure 4 shows the same visual representation of the density distribution within the channel edge facing Prometheus. At these locations density enhancements for both models is generally over a larger area with an inhomogeneous distribution of clumps within. Here, we see a difference in the most prominent clumps, a difference of 7.1% for the average number density and 11.5% for the maximum number density between the two models, fig 5a.

Overall the area showing an enhanced density is larger at this channel edge but when investigated further individual clumps within the enhanced region display increases much greater than the previous edge. This spatial distribution of a larger number of clumps seen over a larger area could account for the reason why we seen an overall higher average surface brightness on the channel edge facing Prometheus in images by CASSINI.

Ultimately these self-gravitating clumps could be dependent on local conditions growing rapidly from one another and might explain why fan structures and moonlets are witnessed predominantly at this channel edge. Additionally the most striking feature witnessed at this channel edge is that the fluctuations in maximum and average density of clumps in the interacting model are chaotic. Here the local density is capable of drastic increases over just a few hours, but also the same is true for the rapid dispersion of clumps, thus suggesting a more unstable environment than the opposing channel edge.

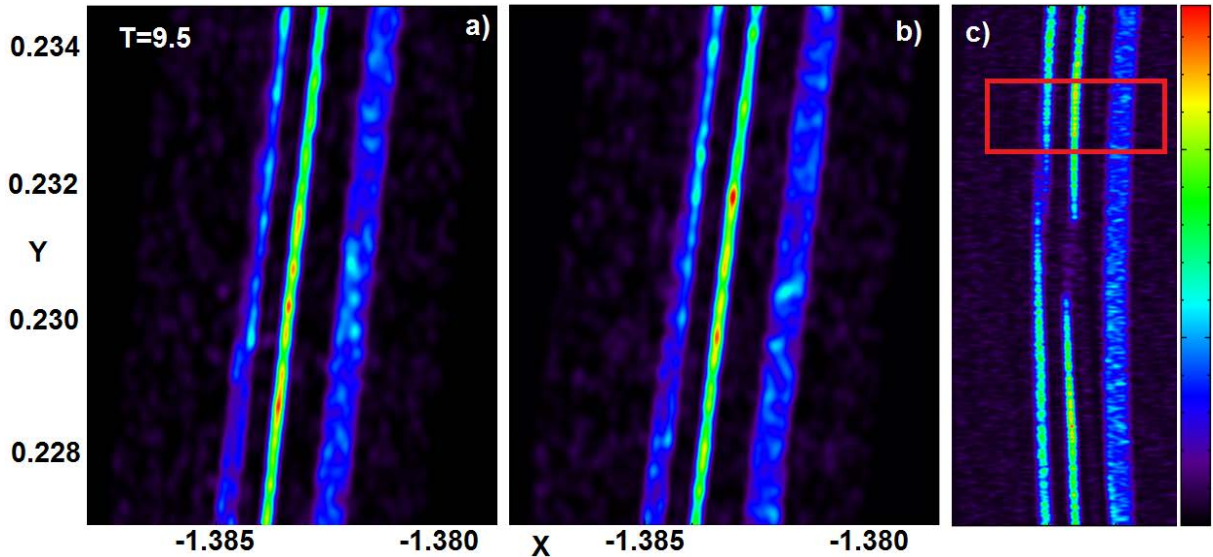

**Figure 4 |** A comparison between the two models is shown at a time of T=9.5, where T is given in Prometheus orbital periods since the start of the simulation and **a)** is the non-interacting model, **b)** is the interacting model **c)** is a visual representation using the interacting model to show the position on the channel that the snapshots **a** and **b** were taken from. Here the reference frame is zoomed in to show where the clump with the highest density is chosen on the channel edge facing Prometheus. The most notable thing here is that unlike in Fig 3, the structure of the opposite channel edge, in the framework of the interacting model does not show a single coherent area of enhanced density. Instead, like in the non-interacting model, the local regions of enhanced density are sprawled here over a larger area with multiple clumps.

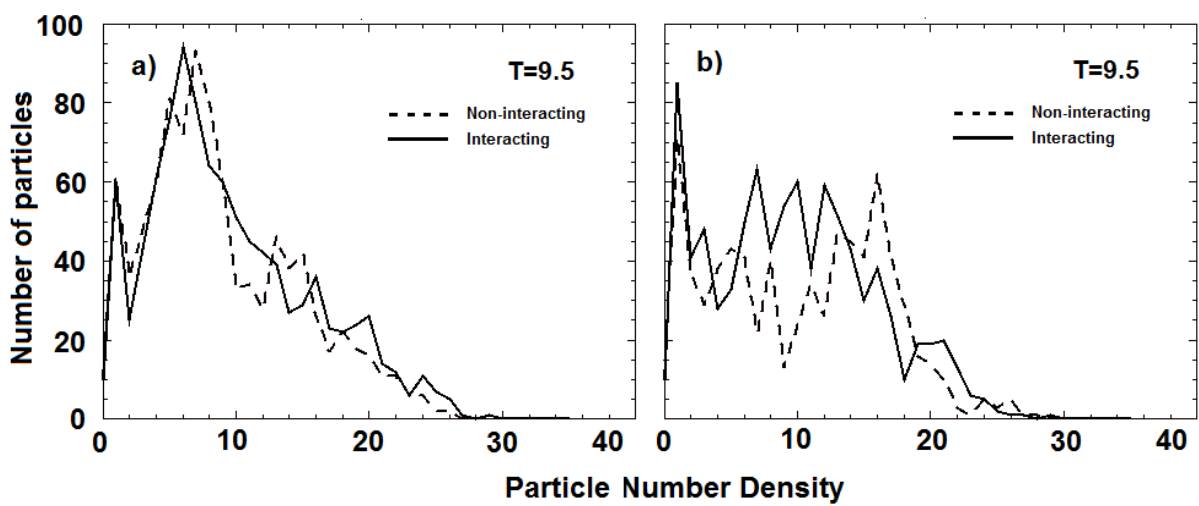

**Figure 5 |** Comparison of density profiles for both the non-interacting and interacting model centred on the most prominent clump where a) is the channel edge facing Prometheus and b) is the channel edge facing away from Prometheus (fig 4). T is the time given in Prometheus orbital periods. The density profiles here are for the most prominent clumps shown in Fig 3 and 4. The profiles of the density distribution in both clumps show that Max No. density does differ from each model, with the most notable being the channel edge facing away from Prometheus (b).

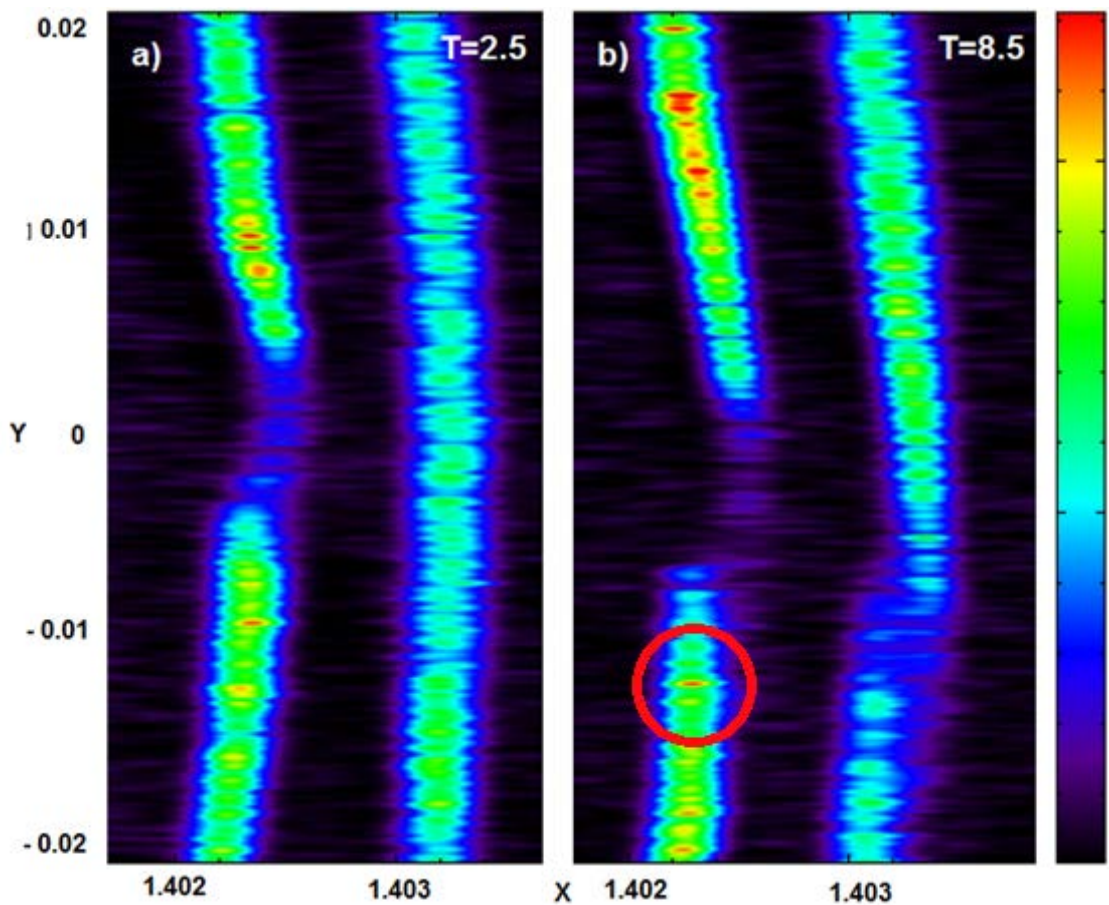

**Figure 6 |** Shown here are the density maps of two channels in the central core (left) obtained from our results, differing in evolution from a) 2.5 orbital periods and b) 8.5 orbital periods from start of simulation for the interacting model only. Both channels are rotated about the origin so that they are centered in the Y axis is zero. The clump circled in b) on the bottom channel edge show considerably higher maximum and average densities than the oppose channel edge and the non-interacting model, see figure 7.

Figure 6 shows the evolution of a channel from 2.5 to 8.5 orbital periods of the interacting model. The most notable thing is the change in locations of highest densities and growth of density, predominantly on the top channel edge (facing Prometheus). This same edge facing Prometheus shows a drifting of the area displaying a density enhancement along with the overall increase in the area. When we consider the density in the largest of the clumps seen at these channel edges, Fig 7 and 8, we see that both channel edges have shown a substantial increase in their average and maximum densities. However at 8.5 orbital periods the clump seen at channel edge facing Prometheus (Fig 8 a) shows a large spike in density compared with the non-interacting model at the same time on the opposing channel edge (Fig 8 b). The exact location of this clump and dramatic local increase is circled in fig 6 b, it should be noted that even the rendering of our model is not able to visualise the extent of this localised increase. This sudden increase in density obtained with the interacting model is short lived decreasing sharply after another orbital period, Fig 5 a. The nature of the chaotic fluctuations can easily be seen when comparing the density profiles seen figure 5a and figure 8a. The overall area where clumps are distributed on the opposing channel edge, bottom Fig 6, does not increase drastically in the same way as the above channel. Moreover there does not appear to be any particle drift as a result of elastic collisions. Here we would expect a diffusion of particles from high density regions to lower density high density areas. From results obtained from our model, Fig 6, we can conclude that low density regions at 2.5 orbital periods continue to decrease while high density regions increase, and in some cases quite dramatically. The low particle density in our model, 0.17 per $km^2$ means that very few collisions would actually take place anyway.

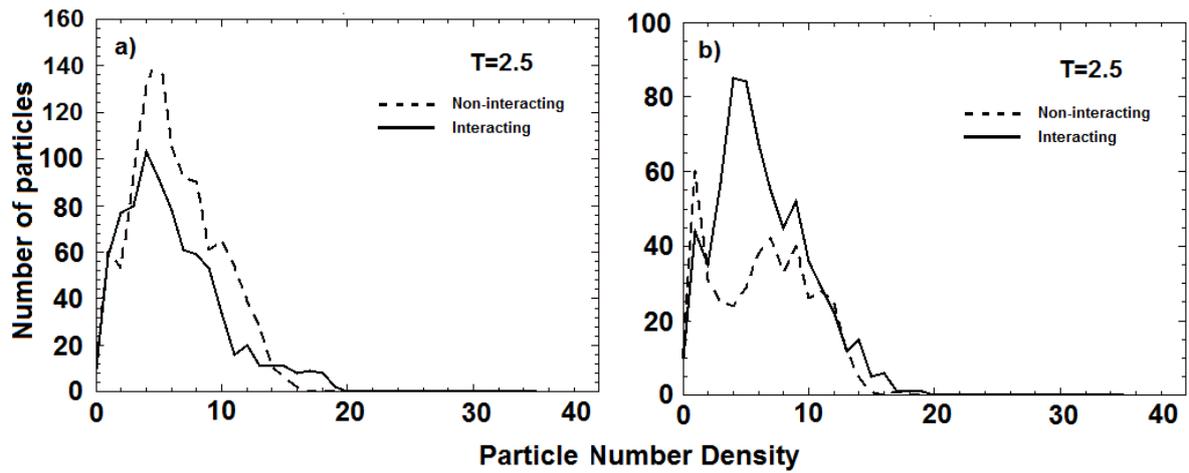

**Figure 7 |** Comparison of density profiles at the channel edges obtained with the non-interacting and interacting models where a) is the channel edge facing Prometheus and b) is the channel edge facing away and T is in Prometheus orbital periods.

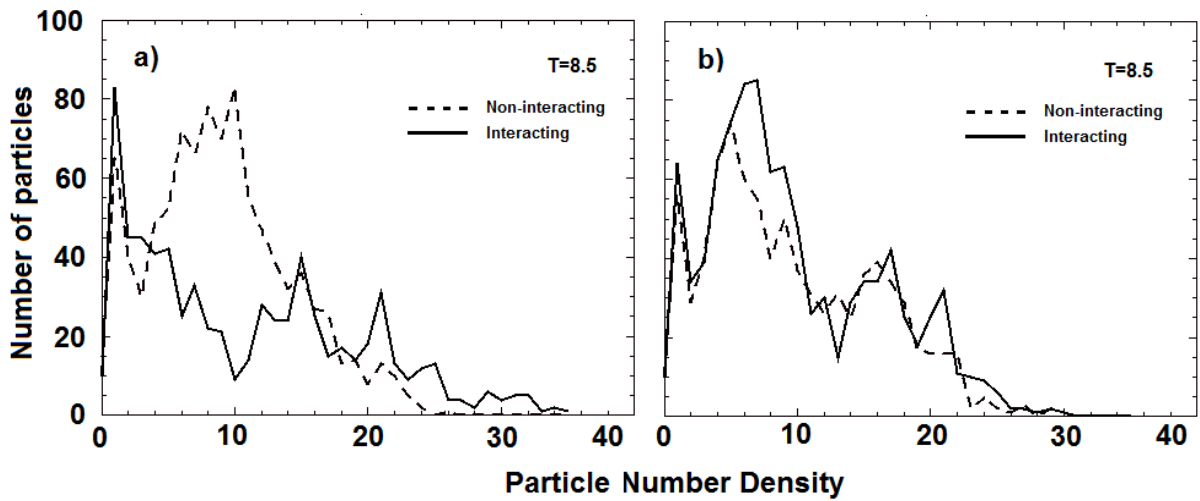

**Figure 8 |** Comparison of density profiles at the channel edges obtained with the non-interacting and interacting models where a) is the channel edge facing Prometheus and b) is the channel edge facing away and T is in Prometheus orbital periods.

## 3.3    Accelerated Growth

Additional analysis of the results obtained with our numerical models was done in the form of mapping, in high resolution, velocity dispersion during an encounter of Prometheus on the F ring. Here two-dimensional rendered plots were created to spatially resolve the variations in the velocity magnitude of ring particles from their initial unperturbed velocity, referred to as $v_d$ velocity dispersion.

$$v_{d\ (t_1-t_2,\ x,\ y,z)} = \left| \vec{v}_{(t_1,\ x,\ y,\ z)} - \vec{v}_{up\ (t_2,\ x,\ y,\ z)} \right| \tag{22}$$

Where $t_1$ and $t_2$ represent a time pre and post encounter of Prometheus on the ring respectively and $v_d$, $\vec{v}$ and $\vec{v}_{up}$ is the magnitude of the velocity dispersion, the velocity post Prometheus encounter and the unperturbed velocity respectively at any given time. Maps were created for the results of both models in the same way that density maps were created in the sense that all particles have a calculated quantity assigned to it, which can then be two-dimensionally rendered. Although here a different colour scheme scale was used to show dispersion values higher (red) and lower (blue), with the unperturbed ring particles set at zero (green). Creating maps like these allowed us to form a link between maximum and minimum velocity dispersions and their locations within the streamer-channel structures. What we find is that a quick glance shows there to be little difference between the results of the two models. However when certain features are zoomed into it starts to become clear that the localised density variations corresponds to the same difference and local variations in velocity dispersion. Figure 9 compares the results of the two models at time T=6.91(Prometheus orbital periods) after the start of the simulation. Here we can see that area of the first two channels (top zoomed into section, fig 9 a and b) show a complete opposite distribution to one another internally. Maximum velocity dispersion calculated in both models at this location is at opposite ends of the feature. The zoomed areas at this time period represent the centre of the channel when Prometheus is at periapsis and where particles rush back into fill the gap. However when we consider the two models there is an asymmetry in the locations of maximum dispersion, the non-interacting (Fig 9 a) shows the highest towards the channel edge facing away (top of zoomed section) while the interacting model displays the most extreme near the edge facing towards Prometheus (bottom of zoomed sections). Another intriguing difference between the two models is that in the framework of the non-interacting model multiple locations of maximum dispersion are observed chaotically positioned within the channel. In contrast the interacting model at the same time period always shows one quite defined area on one channel edge facing towards Prometheus. Density maps of the same area have revealed that there is an asymmetry between the distributions on the channel edges. Again we see a similar asymmetry for the velocity dispersion maps and that this could be an underlying clue to this mystery.

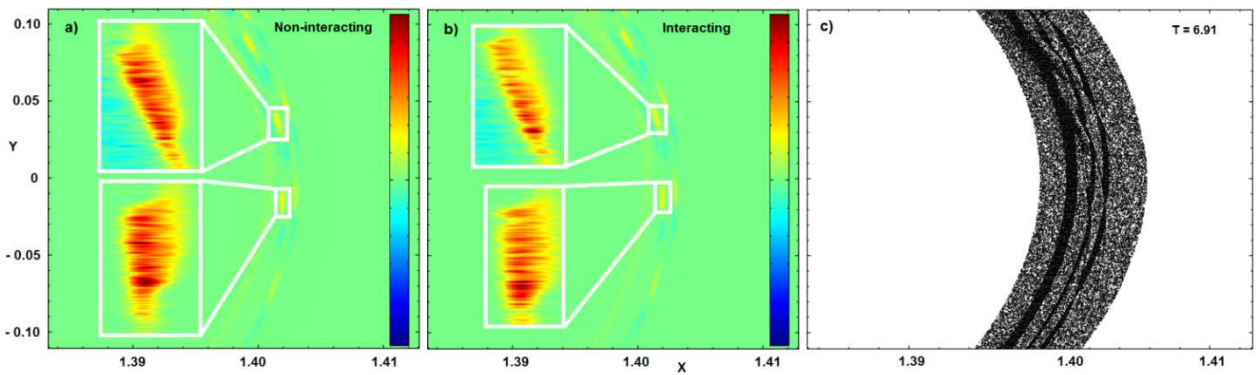

**Figure 9 |** Velocity dispersion maps are shown for both the a) non-interacting model and b) interacting model at a time of T=6.91, where T is given in Prometheus orbital periods and c) shows the positions of particles at the same time. Both the X and Y axis scales are in $2 \times 10^5$ km. The zoomed in section is rescaled to represent the positions of the maximum localised velocity dispersion.

As previously suspected, channel edges facing Prometheus saw local density distributions to be scattered over a larger area with larger fluctuations for the interacting model than the non-interacting model. The possible culprit of this was suggested to be additional turbulence in the movement of particles at this location. The non-interacting model repeatedly produces a more chaotic distribution of particles exhibiting the maximum dispersion consistent with the more dispersed distribution of localised density enhancements on both channel edges.

If the numerical simulations are allowed to evolve further, say, another two orbital periods, we see deterioration in the asymmetry in the local distribution of velocity dispersion between the models (fig 10). Although there does still exist some asymmetry between the areas showing the maximum values it has somewhat diminished. The interacting model still shows a bias towards the channel edge facing Prometheus and the non-interacting model on the edge facing away.

However it is the area around the centre of the channel that now shows the divergence between the results obtained with the two models. Here on the channel edge after 7.5 orbital periods of the initial encounter (Figure 10 a) i) there are individual particles that show negative velocity dispersion amongst the area that has overall increased velocity dispersion. This shows the main differences associated with the gravitational interactions taken into account in one of the models. Particles that move on their own trajectories are free to evolve through Keplerian shear with no additional forces outside those of Saturn and Prometheus; it is these particles that are starting to bleed into the streamer-channel structures creating an inhomogeneous distribution of velocity dispersion.

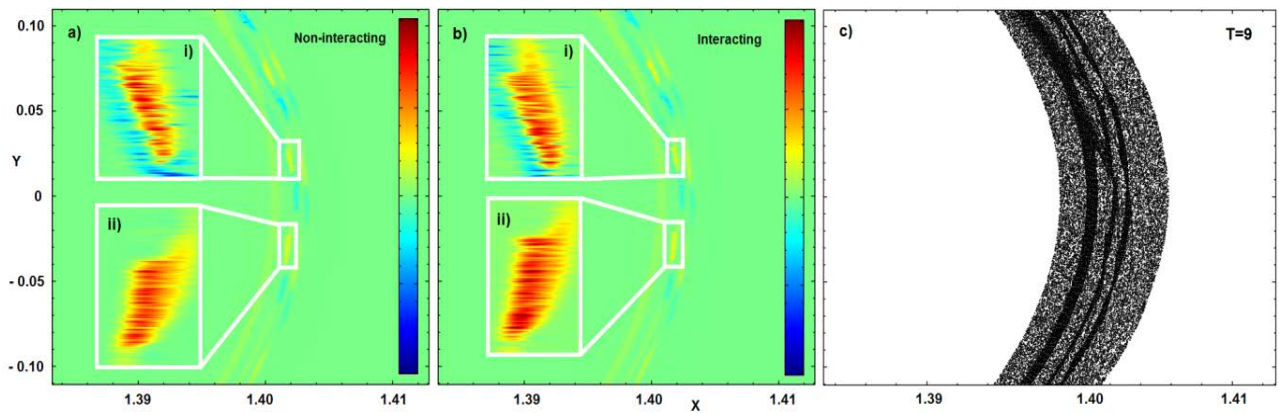

**Figure 10 |** Velocity dispersion maps are shown obtained in the framework both the a) non-interacting model and b) interacting model and c) shows the positions of particles at a time of T=9, where T is given in Prometheus orbital periods. Both the X and Y axis scales are in 2 x $10^5$ km. The zoomed in section are rescaled to represent the positions of the maximum localised velocity dispersion, where i)

and ii) is the centre of the channels position when Prometheus is at periapsis of its orbit 7.5 and 8.5 orbital periods after the initial encounter respectively.

If we now look at the younger structures created by the encounter, 0.5 orbital phases different than those presented in figs 9 and 10, there appears to be little differences between the results of the two models. The most notable difference is shown in fig 11 where the channels are at an orbital phase corresponding to when the channels are close to their most open position. Here, on the channel created after two orbital periods from the initial Prometheus encounter (zoomed section, fig11) we see a difference only in the particles displaying an increase in their velocity magnitude compared (orange) with the normal Keplerian velocity. The distribution of particles with a lower velocity shows no difference between the two models close to the encounter. What we find is that the non-interacting model provides a more dispersed velocity distribution and a much larger spike in the local velocity than the interacting model for the same chosen area. We obtain that these younger areas in the channels (inner strand) quickly become less important to the evolution of the system and the ultimate distribution of localised density enhancements because Keplerian shear promptly distorts and mixes particles in this area.

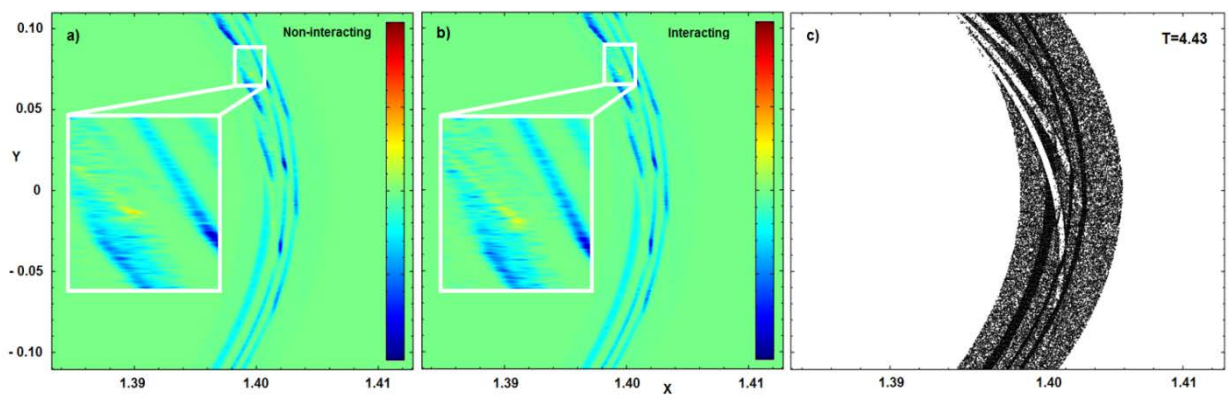

**Figure 11 | |**Here we show velocity dispersion maps obtained with the a) non-interacting model and b) interacting model at a time of T=4.43, where T is given in Prometheus orbital periods and c) shows the positions of particles at the same time. Both the X and Y axis scales are in 2 x $10^5$ km. The zoomed in sections show areas obtained with the two models that display differences in the

distribution in the local velocity dispersion, at the time when the channels are near to their most open position. The zoomed in section showing the difference between the results of two models is on a channel created two orbital periods after initial encounter towards the edge facing Prometheus.

To further investigate the idea of a spatial link between maximum densities and velocity dispersions direct comparisons were made in Figure 12. Here we have identified the locations of maximum density and dispersion and found generally that there is a spatial link between the highest density and the highest dispersions obtained with the interacting model. In contrast the non-interacting model does not explicitly show the same link with a disparity in the locations of maximum density and dispersion.

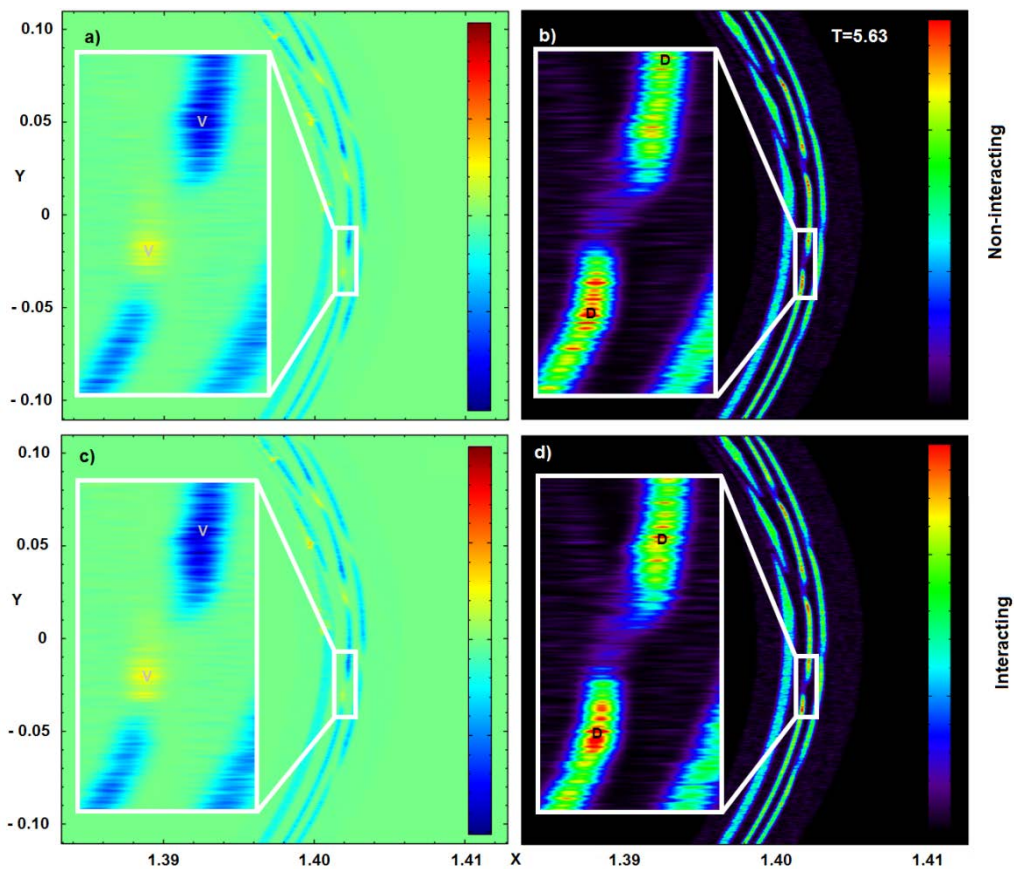

**Figure 12 |** Velocity dispersion and density maps are shown for both the a), b) non-interacting model and c), d) interacting model at a time of T=5.63, where T is given in Prometheus orbital periods and Both the X and Y axis scales are in $2 \times 10^5$ km. The zoomed in sections represent a complete channel showing the asymmetry in velocity dispersion and density between the opposing edges. The locations

of the maximum and minimum values for the velocity dispersion and number density are indicated in each zoomed section as V or D respectively. The linear scales used to display both the velocity dispersion and the density in the zoomed sections are rescaled to show the distribution of the maximum values, however both zoomed areas use the same scale and are therefore comparable.

## 4.   DISCUSSION

A spatial investigation of density enhancements in Prometheus induced structures leads to the conclusion that asymmetry between the two channel edges matches the same asymmetry in the real F ring. Here previous images taken by Cassini (Fig 13, Beurle et al 2010) appear to show fan structures predominately on the channel edges facing towards Prometheus. Many of the fans also show a linear growth proportional to orbital period after initial encounter. In our models we see on the same channel edge the greatest sudden increases in local density after 5 orbital periods in the interacting model. Where density increases beyond the Roche density for F ring clumps could become coherent objects capable of forming fans over many subsequent orbits. The spatial agreement in our models and the real F ring suggest that the large fluctuations seen at this channel edge could be responsible for the embedded objects that create the fans on this particular channel edge. Or if not solely the outcome of a single encounter on a homogeneously distributed core, as we have modelled, the density increases on an already chaotically distributed core by Prometheus would lead to the collapse of clumps on a random selection of channel edges. Again a random distribution of fans is seen on channel edges with many not showing any evidence of embedded moonlets.

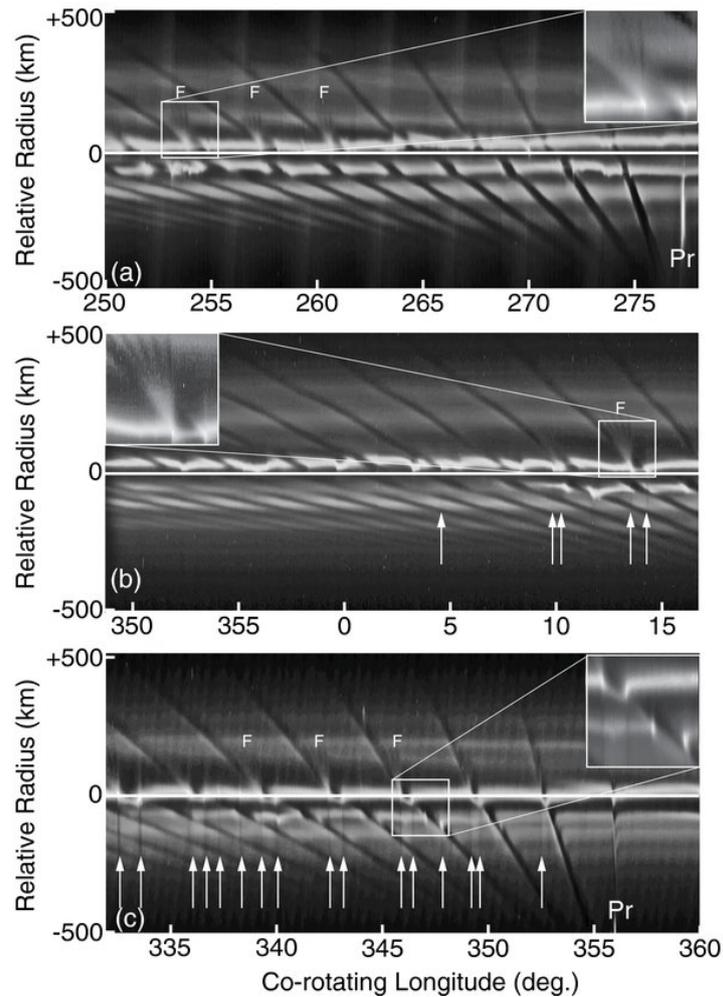

**Figure 13 |** Taken from Beurle et al 2010 the figure shows three 28° sections of the F ring created by mosaics of multiple images where the x-axis represents a co-rotating longitude system and the y-axis as a relative radial position from Saturn. In all frames "F" indicates the longitudinal position of an embedded object in the central core and where appropriate "Pr" denotes the position of Prometheus. Time of image capture is (a) 2008 July 5, (b) 2008 December 8 and (c) 2009 April 16.

A surprising outcome from our simulations was the discovery of a spatial link to the highest velocity dispersions and the highest density enhancements, found only in the framework of the interacting model. General consensus surrounding velocity dispersion and maximum number densities of particles is that high velocity dispersions lead to a lowering of the local density. Amounting to a fragmentation of the area and has been shown to be consistent across many different models of Saturn's rings (Beurle et al 2010; Esposito et al 2012; Torres, Madhusudhanan & Esposito 2013). In our simulations the highest velocity dispersions and densities occur at the same time and location. However, this result could also give light to the reason why we see large chaotic fluctuations in density at the channel edge facing Prometheus. The high velocity dispersions, located on one channel edge give rise to a more turbulent and unstable environment that the clumps reside in. However, it also could in some cases help in the rapid collapse to form coherent objects or moonlets and the apparent random positioning of fans on some channel edges.

## 4.1 Considerations when employing our models

If we consider that the viscosity component may originate from the collisions and gravitational interaction between particles we can make some assumptions and estimate the magnitude and value of the likely viscosity present. The collision rate between particles will be an important factor influencing the outcome of the viscosity magnitude. In our simulations we have an initial maximum average particle density of 0.17 particles per $km^2$. Low particle densities such as these are unlikely to experience any significant rate of collisions, especially when collisions rates are proportional to the distance that separates the particles. Even when physical particle sizes are considered due to the small particle density considered the collision rates are unlikely to rise to anything significant. Thus, for the results obtained with our models physical collisions between particles have negligible effect on the overall evolution of the system, regardless of the collisionless / collisional dynamics employed. Unlike other modelling that investigated the effect of larger stationary moonlets in the A ring (Lewis & Stewart 2009), with which the particles would collide, we are only concerned with gravitational interactions between ring particles. Here, Prometheus is assumed to have very little physical collisions with F ring particles, instead the evolution of the system is dominated by the effect of gravitational scattering. The smaller particle sizes known to exist in the F ring (Scharringhausen & Nicholson 2013) also mean that the magnitude of any likely viscosity resulting from collisions would be much smaller than witnessed in Saturn's main rings.

As particle number density is proportional to collision rates, where particle sizes remain the same: a high density relates to a high viscosity and a low density to a lower viscosity. Fluids comprised of particles with larger radius will consequently expect to have higher collisional rates than fluids with the same number density but with particles of a smaller radius. If we do assume particles to have a physical size, collisions will occur when the particles are moving towards each other as well as when the shortest distance between them on their trajectories will be less than the sum of their radii. To make this assumption all particles should be treated as solid spheres and of the same radius. For collisions to happen the smallest separation between them will be $2a$, where $a$ is the radius of a particle. Therefore the collisional cross section can be shown as: $A = 4\pi a^2$. From this we can see that the area where collisions can occur increases as a square with increasing particle radius $a$. The probability of a collision will increase proportional to the cross sectional area so that the probability $\propto a^2$. With particles that are very small and treated as point masses (as is the case with in our models) the chance of collisions is very small while the collisional probability for much larger particles is considerably larger. It seems that this collisional component of viscosity is unlikely to contribute anything of significance to the evolution of our system.

The other factor, which is a stronger contribution effecting viscosity, is the gravitational attractive forces between particles. Here, like for all other astrophysical discs we have a shearing flow as a function of $r$, which is the radius from the orbital centre (Saturn) of the system. A velocity gradient exists between Keplerian shearing flow layers as function of $r$, where the velocity of the layer is proportional to $\sqrt{r}$. When two layers are located at $r_1$ and $r_2$ the difference in the Keplerian flow can be given as:

$$v_s = \sqrt{\frac{\mu}{r_1}} - \sqrt{\frac{\mu}{r_2}} \qquad (23)$$

Here we assume circular orbits. The total force exerted by one layer onto another assuming fluid mechanics of a gas can be shown as:

$$F = k(2\pi r_1 . z) . \left( \frac{\left(\sqrt{\frac{\mu}{r_1}} - \sqrt{\frac{\mu}{r_2}}\right)}{(r_2 - r_2)} \right) \quad (24)$$

Where $r_1$ and $r_2$ is the radial location of these two layers, $z$ is the vertical height of the disk, $(r_2 - r_2)$ is the separation between the layers and $(2\pi r_1 . z)$ is the area under consideration.

The viscous shear stress can be expressed as $\tau = \frac{F}{A} = k \left( \frac{\left(\sqrt{\frac{\mu}{r_1}} - \sqrt{\frac{\mu}{r_2}}\right)}{(r_2 - r_2)} \right)$ \quad (25)

After simple calculations, considering thinner layers it is further reduced to $\tau = k \left( \frac{\sqrt{\mu^{1/2}}}{R^{3/2}} \right)$, where $R$ is the radial distance to the layer under consideration.

From the above equation (24) it is clear to see that the vertical height, $z$, of the astrophysical disc in question plays an important role in the magnitude of the viscosity. Our models only assume a monolayer of particles which again reduces the overall force exerted on each layer.

Another element that was ignored and deemed to have a negligible effect on the evolution of the system for the time scales being observed was the mutual precession between the ring particles and Prometheus' orbits. At 0.057°d$^{-1}$ the precession between the orbits and their alignment varies very little when only considering a few orbital periods (Chavez 2009).

Although it is true that a more complete system of Saturn's moons will have an effect on vortex formation in any of the rings we believe that this will have a negligible effect on the very short term effects of localised disruption during an encounter. Since we are initially looking, in this manuscript, to investigate density asymmetry created during the encounter with Prometheus and not the long term effects built up with resonances over time we feel this is a fair assumption to make. It is likely that the most significant effects of additional moons will be seen when Prometheus or the disrupted F ring material is in a radial alignment with Saturn and the additional moons. As a result it is likely that during the time interval of our simulations this would not be seen. However this question does pose some interest to our models. Modelling the extended satellite system of Saturn in additional to our current simulations would prove to further clarify any impact of external moons on the evolution of F ring particles.

**Acknowledgements**

This research has made use of data obtained by the CASSINI project from the Planetary Data System (PDS).

**Author contributions**

P. J. S performed the numerical simulations, analysed the results and wrote the manuscript.
F. V. K contributed ideas, took part in discussions of results and helped with editing of the manuscript.

**Competing financial interests**

The authors declare no competing financial interests.